\documentclass[useAMS,usenatbib]{mn2e}
\usepackage{graphicx}
\usepackage{amsmath}
\usepackage{array}
\usepackage{multicol}
\usepackage{multirow}
\usepackage{lscape}
\usepackage{rotating}
\usepackage{url}
\usepackage{color}
\usepackage[colorlinks=true,
            linkcolor=blue,
            urlcolor=blue,
            citecolor=black]{hyperref}

\title[Self-consistent magnetar model]{A self-consistent analytical magnetar model: The luminosity of $\gamma$-ray burst supernovae is powered by radioactivity}

\author[Cano et al.]{\noindent Zach Cano$^{1,2}$\thanks{zewcano@gmail.com}, Andreas K. G. Johansson$^{1}$ \& Keiichi Maeda$^{3,4}$  \\
\noindent $^{1}$Centre for Astrophysics and Cosmology, Science Institute, University of Iceland, Dunhagi 5, 107 Reykjavik, Iceland.\\
\noindent $^{2}$International Research Fellow of the Japan Society for the Promotion of Science.\\
\noindent $^{3}$Department of Astronomy, Kyoto University, Kitashirakawa-Oiwake-cho, Sakyo-ku, Kyoto 606-8502, Japan.\\
\noindent $^{4}$Kavli Institute for the Physics and Mathematics of the Universe (WPI), University of Tokyo, 5-1-5 Kashiwanoha, Kashiwa, \\
Chiba 277-8583, Japan. \\
}

\begin{document}

\date{Accepted xx. Received xx; in original form xx}

\pagerange{\pageref{firstpage}--\pageref{lastpage}} \pubyear{2013}

\maketitle

\label{firstpage}

\begin{abstract}
We present an analytical model that considers energy arising from a magnetar central engine. The results of fitting this model to the optical and X-ray light curves (LCs) of five long-duration $\gamma$-ray bursts (LGRBs) and two ultra-long GRBs (ULGRBs), including their associated supernovae (SNe), show that emission from a magnetar central engine cannot be solely responsible for powering an LGRB-SN.  While the early AG-dominated phase can be well described with our model, the predicted SN luminosity is underluminous by a factor of $3-17$.  We use this as compelling evidence that additional sources of heating must be present to power an LGRB-SN, which we argue must be radioactive heating.  Our self-consistent modelling approach was able to successfully describe all phases of ULGRB~111209A / SN~2011kl, from the early afterglow to the later SN, where we determined for the magnetar central engine a magnetic field strength of $1.1-1.3\times10^{15}$~G, an initial spin period of $11.5-13.0$~ms, a spin-down time of $4.8-6.5$~d, and an initial energy of $1.2-1.6\times10^{50}$~erg.  These values are entirely consistent with those determined by other authors.  The luminosity of a magnetar-powered SN is directly related to how long the central engine is active, where central engines with longer durations give rise to brighter SNe.  The spin-down timescales of superluminous supernovae (SLSNe) are of order months to years, which provides a natural explanation as to why SN~2011kl was less luminous than SLSNe that are also powered by emission from magnetar central engines.
\end{abstract}

\begin{keywords}
xxx
\end{keywords}

\section{Introduction}

In the current paradigm of $\gamma$-ray burst (GRB) phenomenology, there are two favoured theoretical models that explain the physics of the central engine powering these events: the collapsar model (Woosley 1993; MacFadyen \& Woosley 1999; MacFadyen et al. 2001; Zhang et al. 2004), and the millisecond magnetar model (Usov 1992; Duncan \& Thompson 1992; Wheeler et al. 2000; Thompson et al. 2004; Metzger et al. 2015, M15 hereafter).  

In the collapsar model, a disk forms around a black hole (BH), and the accretion of stellar material onto the compact object leads to the ejection of shells that are collimated in a jet, which is either driven by neutrino annihilation (e.g. MacFadyen \& Woosley 1999) or activated by the Blandford--Znajek mechanism (e.g. Komissarov \& Barkov 2009), at relativistic velocities.  Multiple shells interact producing the initial $\gamma$-ray pulse (the prompt emission), and as they propagate away from the explosion they encounter circumstellar material, producing a long-lived afterglow (AG).  In the millisecond magnetar model (simply referred to as the magnetar model, or just magnetars, in this paper), energy extracted from the rapidly rotating, magnetised NS is collimated into a narrow jet along the object's rotational axis, which then powers the prompt emission and the resulting AG.  In both models the observed radiation is synchrotron, and possibly inverse Compton, in origin.

A basic assertion of the collapsar model is that the duration of the GRB prompt emission is the difference between the time that the central engine operates (i.e. $T_{90}$; though see Zhang et al. (2014) who argue that $T_{90}$ is not a reliable indicator of the engine activity timescale) minus the time it takes for the jet to breakout of the star: $T_{90} \sim t_{\rm engine}$ -- $t_{\rm breakout}$.  A direct consequence of this premise is that there should be a plateau in the distribution of $T_{90}$ for GRBs produced by collapsars when $T_{90}$ $<$ $t_{\rm breakout}$, which was confirmed by Bromberg et al. (2012).  Moreover, the value of $T_{90}$ found at the upper-limit of the plateau seen in three satellites (BATSE, \emph{Swift} and FERMI) was approximately the same ($T_{90}\sim$ 20--30 s), which can be interpreted as the typical breakout time of the jet.  This short breakout time suggests that the progenitor star at the time of explosion is very compact ($\sim 5$ $R_{\odot}$; Piran et al. 2013).

Observational evidence has also been proposed for GRBs, or episodes occurring during a GRB event, being powered by a magnetar central engine.  Using the analytical model derived by Zhang \& M\'esz\'aros (2001, ZM01 hereafter), Rowlinson et al. (2013) modelled the X-ray AGs of a large sample of short-duration GRBs (SGRBs), and in doing so extracted estimates of the magnetic field strengths and initial spin periods of the assumed magnetar central engines.  The presence of a magnetar central engine has been invoked for several long-duration GRBs (LGRBs), including GRB~030227 (Watson et al. 2003), 060729 (Xu et al. 2009), 130215A (Cano et al. 2014), 130427A (Bernardini et al. 2014) and 130831A (De Pasquale et al. 2015b).  In these models, it is the AG-dominated phase that is usually modelled, such as the plateaus seen for GRBs 060729 and 130215A.  For GRB~130831A, the sharp decline seen around 80~ks was interpreted as the turn-off of an energy-injection episode, which was supposed as energy injection from a magnetar by De Pasquale et al. (2015b).  Watson et al. (2003) provided arguments that the only solution to explain the X-ray line emission seen for GRB~030227 was some form of directed pulsar emission, which collimated a large amount of energy into the remnant for several thousand seconds after the prompt emission.

To date, determining whether LGRBs are produced by accreting BHs or magnetars has been hotly debated.  Original calculations suggested that the maximum amount of energy that could be extracted from a magnetar central engine was of order $2-5\times10^{52}$~erg (e.g. Usov 1992).  Recent investigations by Mazzali et al. (2014) and Cano et al. (2015) demonstrated that the kinetic energy of most, if not all, gamma-ray burst supernovae (GRB-SNe) do not exceed this limit.  However, recent calculations performed by M15 showed that, for a wide range of equation of states, the maximum amount of rotational energy that can be extracted from a magnetar central engine is likely to be closer to $10^{53}$~erg.  In the collapsar model there is no theoretical upper limit to the maximum amount energy available to power a GRB event as the total energy that can be extracted from the accreting BH can be continuously replenished via accretion.  As such, this once-believed discriminant between the two models is likely to be no longer valid, and new means of addressing this long-lasting debate must be concocted.

In an attempt to solve this outstanding problem, we developed an analytical model to determine if emission arising from a magnetar central engine can be responsible for powering all phases of an LGRB event, from the prompt emission, to the AG and finally the SN.  In doing so we have given self-consistent evidence that the emission coming from a given LGRB-SNe cannot be powered solely by energy from a magnetar, and that additional sources of heating are required, which we argue is very likely to be radioactive heating.  We also show that ultra-long GRBs (ULGRBs) and their associated SNe can indeed be powered entirely by spin-down energy extracted from a magnetar central engine, and discuss this result in the context of GRB-SNe and superluminous supernovae (SLSNe; Quimby et al. 2011; Gal-Yam 2012).  

This paper is arranged as follows.  In Section \ref{sec:Model} we present our derived theoretical model, building up previous analytical models that consider energy output from a magnetar central engine.  In Section \ref{sec:method} we describe our method, while in Section \ref{sec:Results} we discuss the results of fitting our model to the optical and X-ray light curves (LCs) of several GRB-SNe and ULGRBs.  We discuss our results in Section \ref{sec:discussion} and present our conclusions in Section \ref{sec:conclusions}.  Throughout the paper we use the convention $f_{\nu,t}(t) \propto t^{-\alpha}\nu^{-\beta}$, where $\alpha$ and $\beta$ are the temporal and spectral indices, respectively.

\section{Analytical Model}
\label{sec:Model}

\subsection{The Afterglow Phase}

The general idea considered here is energy injection arising from a magnetar central engine, which deposits Poynting flux dominated dipole radiation into the ejecta (e.g. ZM01; Dall'Osso et al. 2011).  The general scenario of how EM radiation is emitted by the magnetar central engine\footnote{In this paper, the model described here is in essence a pulsar-type model, i.e. the tapping of rotational energy from a spinning neutron star.  This is in contrast to other models that define a magnetar as an object that produces emission via the dissipation of magnetic-field energy, such as a soft gamma repeater (e.g. Thompson \& Duncan 1995).} (which also applies to the SN-powered phase; Section \ref{sec:mag_SN_theory}) is as follows (Usov 1992): A strong electric ($\vec{E}$) field  is generated near the stellar surface.  EM forces rip electrons ($\rm e^-$) and positrons ($e^+$) from the the surface creating an electron--positron plasma, which is accelerated by the $\vec{E}$-field to ultra-relativistic velocities.  This acceleration leads to the emission of $\gamma$-rays by the primary particles, which then interact with the ultra-strong magnetic ($\vec{B}$) field ($\gamma + \vec{B} \rightarrow \mathrm{e^-} + \mathrm{e^+} + \vec{B}$) to create a secondary electron--positron plasma that emits additional EM radiation, which is either synchrotron and/or Inverse Compton.

Our phenomenological model, which is based on the theoretical framework of ZM01, considers an initial impulsive energy input, and then a continuous energy input.  We made similar assumptions as Rowlinson et al. (2013) by considering a canonical NS with a mass of $1.4$~M$_{\odot}$ and a radius of $10^{6}$~cm, which allowed us to reduce the number of free parameters in the fit.  The analytical form of the energy-injection magnetar model is: 

\begin{equation}
L_{\rm AG}(t)=L_{0}\left(1+\frac{t}{T_{0}}\right)^{-2} \hspace{5pt} ({\rm erg~s^{-1}})
\label{equ:magnetar_AG}
\end{equation}

\noindent where $L_{0}$ is the plateau luminosity, $T_{0}$ is the plateau duration.  During this phase, energy injection from the magnetar is expected to be an additional component to the typical AG LC.  As such we also consider an additional single power-law (SPL) component (e.g. Rowlinson et al. 2013), which is analogous to the impulsive energy input term in the model of ZM01 (see Section \ref{sec:discussion} for further discussion):

\begin{equation}
L_{\rm SPL} (t) = \Lambda t^{-\alpha} \hspace{5pt} ({\rm erg~s^{-1}})
\label{equ:SPL}
\end{equation}

\noindent where $\Lambda$ is the normalisation constant and $\alpha$ is the decay constant.  Here we assume $\alpha = \Gamma_{\gamma} + 1$, where $\Gamma_{\gamma}$ is the photon index of the prompt emission, assuming that the decay slope is governed by the curvature effect (e.g. Kumar \& Panaitescu 2000; Piran 2004).  The photon index for each GRB in our sample can be found in Table \ref{table:best_fit}.

The values of $L_{0}$ and $T_{0}$ can be related back to equations 6 and 8 in ZM01 (see as well Rowlinson et al. 2013) to estimate the magnetic field strength ($B$; eq. \ref{equ:mag_AG_B}) and the initial spin period ($P$; eq. \ref{equ:mag_AG_P}) of the magnetar:

\begin{equation}
B = \sqrt{\frac{4.2}{L_{0,49}T^{2}_{0,-3}}} \hspace{5pt} ({\rm 10^{15}~G})
\label{equ:mag_AG_B}
\end{equation}

and

\begin{equation}
P = \sqrt{\frac{2.05}{L_{0,49}T_{0,3}}} \hspace{5pt} ({\rm ms})
\label{equ:mag_AG_P}
\end{equation}

\noindent where $L_{0,49} = L_{0} / 10^{49}$~erg~s$^{-1}$ and $T_{0,3} = T_{0} / 10^{3}$~s.

\subsection{Magnetar-powered Supernova}
\label{sec:mag_SN_theory}

Next, we consider the analytical model proposed by Kasen \& Bildsten (2010) (originally concocted by Ostriker \& Gunn 1971, and see as well Woosley 2010 and Barkov \& Komissarov 2011), who consider that the LCs of SLSNe can be powered by energy released from the spin-down of a newly formed magnetar.  We utilized the analytical derivation of Chatzopoulos et al. (2011), who use the dipole spin-down formula (their eq. 11) as the energy deposition function (instead of energy supplied by the radioactive decay of nickel into cobalt, and cobalt into iron), as well as the first law of thermodynamics coupled with the diffusion approximation (Arnett 1980; 1982; Valenti et al. 2008; Chatzopoulos et al. 2009) to predict a magnetar-powered SN LC (eq. \ref{equ:mag_SN}).  As in the previous model, the radius of the magnetar is assumed to be 10$^6$ cm (i.e. 10 km), and we considered an $l=2$ magnetic dipole.

\begin{equation}
 L_{\rm SN}(t) = \frac{E_{\rm p}}{t_{\rm p}}~{\rm exp}\left(\frac{-x^2}{2}\right)\int_0^x~\frac{z~{\rm exp}\left(\frac{z^2}{2}\right)}{(1+yz)^2}\, \mathrm{d}z \hspace{10pt} ({\rm erg~s^{-1}})
 \label{equ:mag_SN}
\end{equation}

\noindent where $E_{\rm p}$ is the initial energy of the magnetar (units of erg) and $t_{\rm p}$ is the characteristic spin-down time of the magnetar (units of days). Additionally, $x = t / t_{\rm diff}$ and $y = t_{\rm diff} / t_{\rm p}$, where $t_{\rm diff}$ is the diffusion timescale of the SN in units of days.  The magnetic field (eq. \ref{equ:mag_SN_B}) and initial spin period (eq.~\ref{equ:mag_SN_P}) of the magnetar can then be calculated as:

\begin{equation}
B = \sqrt{\frac{1.3\times10^{2}~P^2}{t_{\rm p, yr}}} \hspace{5pt} ({\rm 10^{15}~G})
\label{equ:mag_SN_B}
\end{equation}

and

\begin{equation}
P = \sqrt{\frac{2\times10^{46}}{E_{\rm p}}} \hspace{5pt} ({\rm ms})
\label{equ:mag_SN_P}
\end{equation}

\noindent where $t_{\rm p, yr}$ is the characteristic spin-down time of the magnetar in units of years.

\subsection{Combined Model}

In this paper we added eqs. (\ref{equ:magnetar_AG}), (\ref{equ:SPL}) and (\ref{equ:mag_SN}) together and fit it to the $R$-band data of five LGRBs and two ULGRBs (see Table \ref{table:references}):

\begin{equation}
 L_{\rm total} (t) = L_{\rm AG} + L_{\rm SN} + L_{\rm SPL}  \hspace{5pt} ({\rm erg~s^{-1}})
 \label{equ:mag_combined}
\end{equation}

\noindent to determine the initial spin periods and magnetic fields of the magnetar central engine in each event.  However, in order to maintain self-consistency between the two models we recast the free parameters of the SN-powered phase ($E_{\rm p}$ and $t_{\rm p}$) in terms of the free parameters in the AG-powered phase ($L_{0}$ and $T_{0}$).  By considering the respective equations for $P$ and $B$ in the two models (eqs. \ref{equ:mag_AG_B}, \ref{equ:mag_AG_P}, \ref{equ:mag_SN_B} and \ref{equ:mag_SN_P}), we derived:


\begin{equation}
E_{\rm p} = \frac{2.00}{2.05}~L_{0}T_{0} \hspace{5pt} ({\rm erg})
\label{equ:E_LT_link}
\end{equation}

and

\begin{equation}
t_{\rm p} = 2~T_{0}  \hspace{5pt} ({\rm s})
\label{equ:tp_T_link}
\end{equation}

\noindent  We note that ZM01 also have $E_{\rm p} \approx L_{0}T_{0}$ (for various values of $q$ and $\kappa$, which are dimensionless constants in their analytical model).
Thus these relations were substituted into eq. (\ref{equ:mag_SN}), so that it was recast in terms of $L_{\rm 0}$, $T_{\rm 0}$ and $t_{\rm diff}$.  Then $B$ and $P$ can be calculated using eqs. (\ref{equ:mag_AG_B}) and (\ref{equ:mag_AG_P}), respectively.

At this point, an entire GRB-SN event can be considered as being powered solely by a magnetar central engine if the luminosity of the early AG phase, and the subsequent values of $P$ and $B$, then reproduces both the timescale and luminosity of the SN-dominated phase.  However, as we shall see later on, for all events except ULGRB~111209A / SN~2011kl, this is not the case.  As such, we also considered a normalisation factor in the SN-phase of the event to account for additional sources of heating needed to reproduce the luminosity of the SN:

\begin{equation}
 L_{\rm total} (t) = L_{\rm AG} + \Phi L_{\rm SN} + L_{\rm SPL}  \hspace{5pt} ({\rm erg~s^{-1}})
 \label{equ:mag_combined_norm_SN}
\end{equation}

\noindent where $\Phi$ is an additional free-parameter that was fit to the optical LCs.  Therefore, if a GRB-SN bump  has a value of $\Phi \approx 1$, this event can be considered as being powered entirely by EM emission from a magnetar central engine.  Conversely, for all events where $\Phi > 1$, additional sources of heating are needed to explain the luminosity of the SN phase, which based on previous investigations, is likely to the heating from the radioactive decay of nickel and cobalt into their daughter products.

\begin{table}
\centering
\setlength{\tabcolsep}{2.5pt}
\setlength{\extrarowheight}{5pt}
\caption{References to GRB-SNe datasets}
\label{table:references}
\begin{tabular}{cccccc}
\hline
GRB	&	SN	&	$z$	&	$E(B-V)_{\rm fore}$ 	&	$E(B-V)_{\rm rest}$ &	Ref.	\\
\hline											
041006	&	-	&	0.716	&	0.021	&	0.033	&	($1-4$)	\\
050525A	&	2005nc	&	0.606	&	0.082	&	0.097	&	($5-7$)	\\
090618	&	-	&	0.54	&	0.073	&	0.090	&	(8)	\\
091127	&	2009nz	&	0.49	&	0.033	&	0.000	&	($9-13$)	\\
111209A	&	2011kl	&	0.677	&	0.015	&	0.044	&	(14)	\\
121027A	&	-	&	1.773	&	0.017	&	-	&	-	\\
130831A	&	2013fu	&	0.479	&	0.039	&	0.000	&	($15-16$)	\\
\hline
\end{tabular}
\begin{flushleft}
NB: Units of extinction are in magnitudes.\\
NB: Using foreground extinction maps from Schlegel et al. (1998) and Schlafly \& Finkbeiner (2011).\\
\textbf{References}: (1) \cite{Stanek2006}; (2) \cite{Urata2007}; (3) \cite{Misra2005}; (4) \cite{Kann2006}; (5) \cite{DellaValle2006}; (6) \cite{Klotz2005}; (7) \cite{Levesque2010}; (8) \cite{Cano2011a}; (9) \cite{Troja2012}; (10) \cite{Vergani2011}; (11) \cite{Filgas2011}; (12) \cite{Cobb2010}; (13) \cite{Berger2011}; (14) \cite{Greiner2015}; (15) \cite{Cano2014}; (16) \cite{dePasquale2015b}.
\end{flushleft}
\end{table}

\begin{figure*}
 \centering
 \includegraphics[bb=0 0 505 500,keepaspectratio=true]{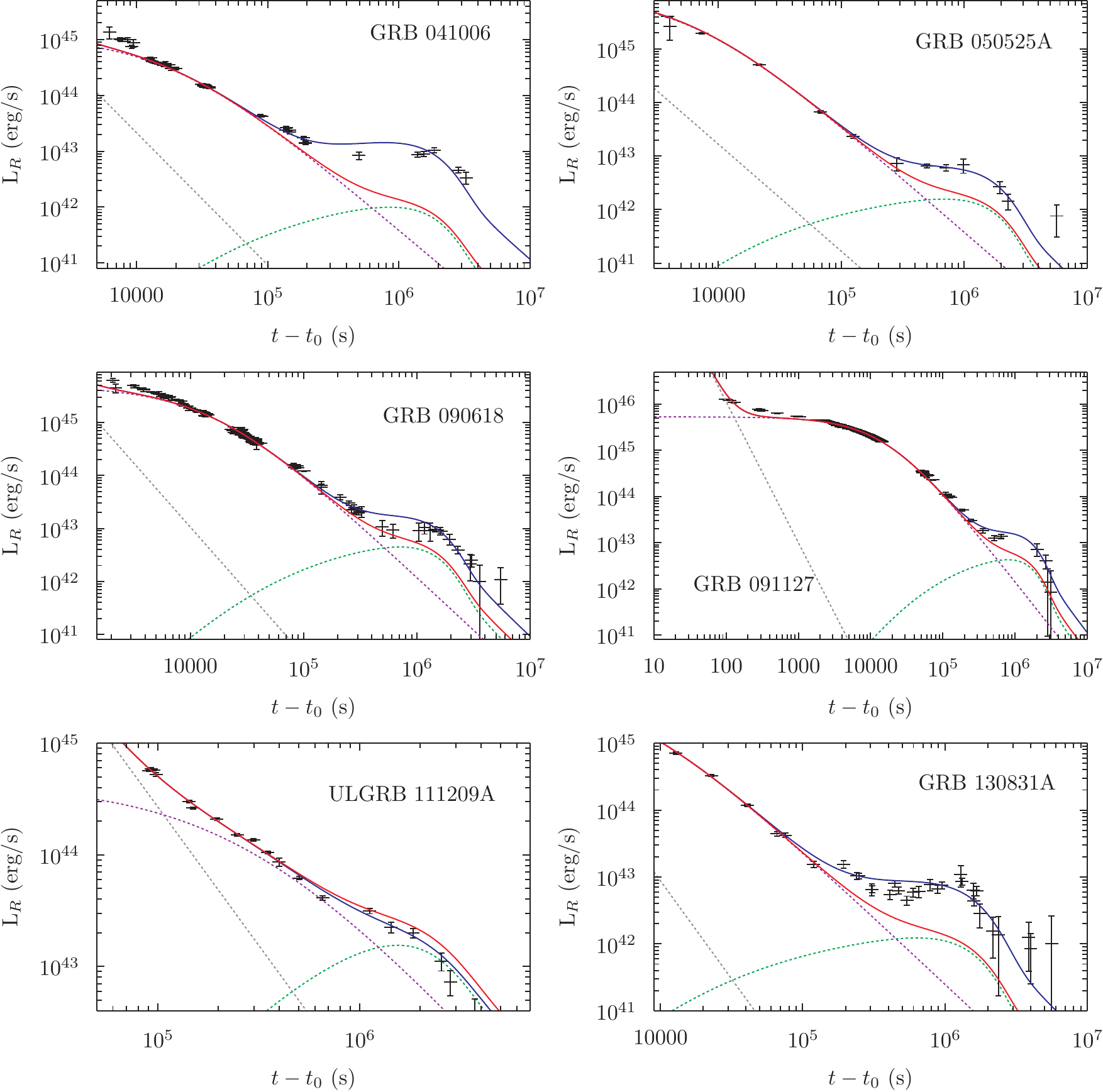}
 \caption{Mosaic of the fitted analytical magnetar model to the $R$-band (observer-frame filter; rest-frame times) LCs of five LGRBs and one ULGRB, and their accompanying SNe: GRB~041006, GRB~050525A / SN~2005nc, GRB~090618, GRB~091127 / SN~2009nz, ULGRB~111209A / SN~2011kl and GRB~130831A / SN~2013fu.  The fitted model has three components: A SPL (grey dashed), energy injection plateau phase (purple dashed) and a magnetar-powered SN (green dashed).  The sum of these three components is shown as the solid red line.  For all events except ULGRB~111209A, while the model (red line) provides a good fit to the early data when the AG dominates, the same model under-predicts the SN luminosity at late times.  For all events we therefore fitted an additional component ($\Phi$; see Section \ref{sec:Model} and Table \ref{table:best_fit}) to get the model to match the observations (blue line).  The fact that the model (red line) under-predicts the SN luminosity is taken as evidence that these SNe are not powered solely by a magnetar central engine, but another source of heating is required, which we argue is radioactive heating.  For ULGRB~111209A, the model (red line) successfully reproduces all phases of the GRB event, and no additional sources of heating are needed to power the accompanying supernova (SN~2011kl).  For ULGRB~111209A, we find a magnetic field of $B=1.1-1.3\times10^{15}$~G, and an initial spin period of $P=11.5-13.0$ ms.  These numbers are derived from the results of fitting both the optical and X-ray (see Section \ref{sec:Results}) data of this event, and are entirely consistent with similar analyses performed by other authors (e.g. Greiner et al. 2015; Metzger et al. 2015).}
 \label{fig:GRB_SN_multiplot}
\end{figure*}

\section{Method \& Analysis}
\label{sec:method}

The LGRBs/ULGRBs in our sample were chosen because they had good coverage of both the AG and SN phases, which would allow this type of modelling to be performed.  Host-subtracted, observer-frame $R$-band observations were dereddened for foreground (Schlegel et al. 1998; Schafly \& Finkbeiner 2011) and rest-frame extinction (see Table \ref{table:references}), and then converted to monochromatic flux densities (mJy) using the zeropoints in Fukugita et al. (1995).  

For GRBs 091127, 111209A and 121027A, we downloaded their XRT (0.3--10 keV) LCs from the UK \emph{Swift}-XRT LC repository\footnote{\url{http://www.swift.ac.uk/xrt_curves/}}.  The X-ray LCs of 050525A (2--10 keV), 090618 and 130831A were taken from Blustin et al. (2005), Cano et al. (2011a) and De Pasquale et al. (2015), respectively.  Note that no X-ray data was obtained for GRB~041006.   The XRT LCs (units of erg~s$^{-1}$~cm$^{-2}$) were converted into monochromatic flux densities (mJy, i.e. erg~s$^{-1}$~cm$^{-2}$~Hz$^{-1}$) using a frequency of $\nu = 1.2\times10^{18}$~Hz, which corresponds to the midpoint of the 0.3--10 keV X-ray passband (i.e. 5 keV).  The flux density X-ray LC was then converted to rest-frame 0.3--10 keV X-ray luminosity LCs using the procedure described in Lamb \& Reichart (2000)\footnote{When calculating the distance luminosities and comoving distances, we used a $\Lambda$CDM cosmology constrained by Planck (Planck Collaboration et al. 2013), where $H_{0} = 67.3$ km s$^{-1}$ Mpc$^{-1}$, $\Omega_{M} = 0.315$, $\Omega_{\Lambda} = 0.685$.}.  In their procedure, knowledge of $\alpha$ and $\beta$ at each moment in time are required.  For GRBs 121027A and 130831A we modelled the X-ray LCs ourselves to determine the decay rates and break times, while for the remaining events we used the values of $\alpha$ determined by various authors.  The precise values of $\alpha$ are described in detail in Section \ref{sec:discussion} for each event.   We then assumed a closure relation of $\alpha = 3/2\beta-1/2$ to calculate $\beta$, which is valid for both ISM (Sari et al. 1998) and wind (Chevalier \& Li 2000) media, and when the observed X-ray flux is always above the cooling frequency (e.g. Gao et al. 2013)

\begin{figure*}
 \centering
 \includegraphics[bb=0 0 503 504,keepaspectratio=true]{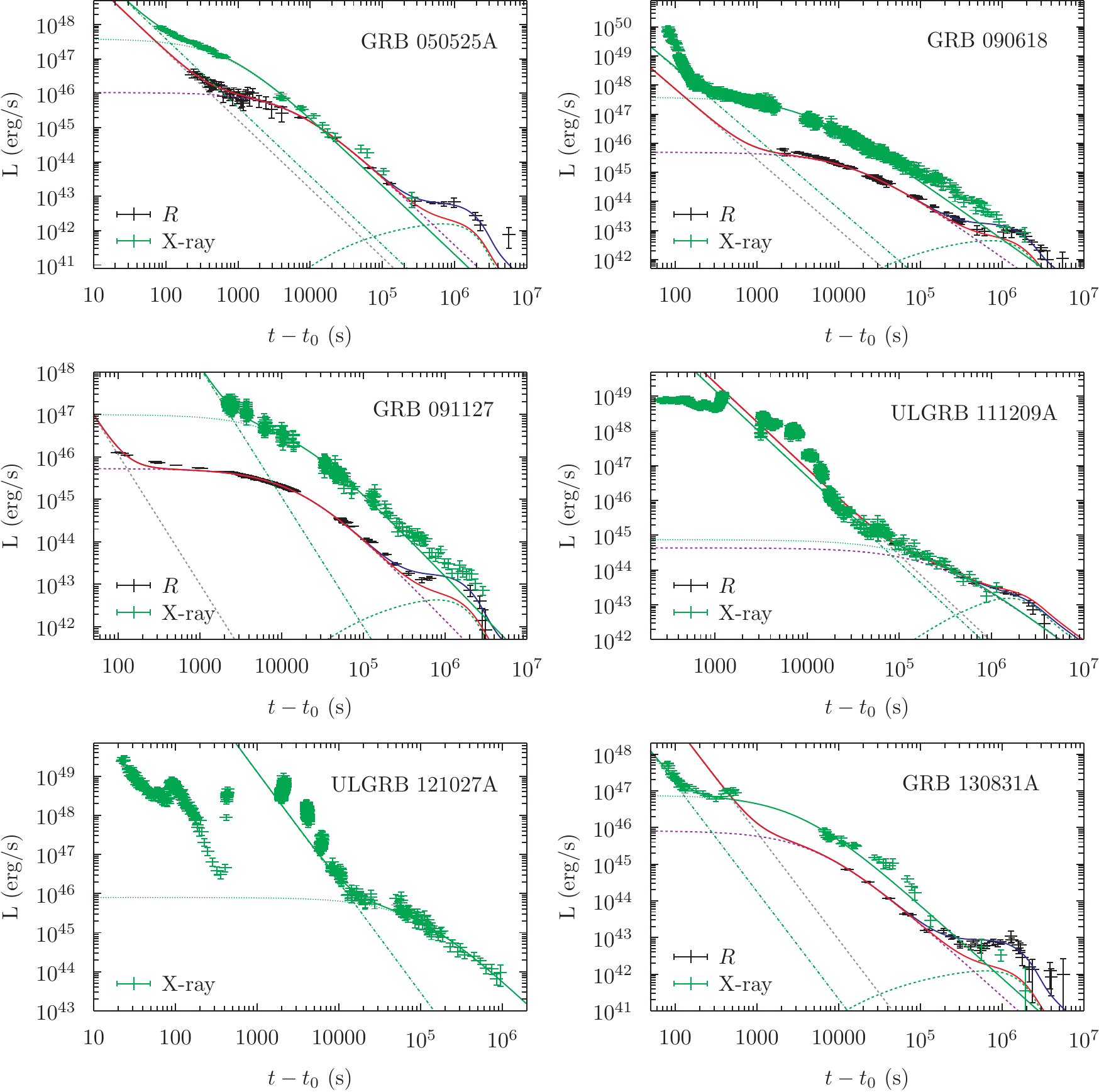}
 \caption{Mosaic of the fitted analytical magnetar model to the X-ray (green points) LCs of six GRBs: GRB~050525A, GRB~090618, GRB~091127, ULGRB~111209A, ULGRB~121027A, and GRB~130831A.  Time is shown in the rest-frame.  The $R$-band LCs (black points) of each event (except ULGRB~121027A) are also plotted for comparison, where the three components are identical to those plotted in Fig. \ref{fig:GRB_SN_multiplot}.  We fit eqs. \ref{equ:magnetar_AG} (green dotted) and \ref{equ:SPL} (green dot-dash), where the sum of both are shown as the solid green line, to the X-ray LCs to determine $L_{0}$, $T_{0}$ and $\Lambda$ for each event.  It is seen that for most events, the values of $T_{0}$ are not consistent between the X-ray and $R$-band data (e.g. GRBs 050525A, 090618 and 130831A).  However for GRB~091127, the values of $T_{0}$ are consistent (although the calculated values of $B$, $P$, $E_{\rm p}$ and $t_{\rm d}$ are not), as too are those seen for ULGRB~111209A, for which the resultant values of $P$, $B$, $E_{\rm p}$ and $t_{\rm d}$ are consistent with those derived from just the $R$-band data.}
 \label{fig:GRB_Xray_multiplot}
\end{figure*}

The (rest-frame) $R$-band luminosity LCs (see Fig. \ref{fig:GRB_SN_multiplot}) were then fitted with the combined model (eq. \ref{equ:mag_combined}) using \textsc{pyxplot}\footnote{\url{http://pyxplot.org.uk}}, where the free parameters (L$_{0}$, T$_{0}$, $\Lambda$, $t_{\rm diff}$, and $\Phi$) were determined via the generic fitting algorithm used by \textsc{pyxplot}, which uses a Bayesian approach to the fitting of models to data.  The (rest-frame) X-ray luminosity LCs (Fig. \ref{fig:GRB_Xray_multiplot}) were fit with eq. \ref{equ:magnetar_AG} and eq. \ref{equ:SPL}, where the free parameters were L$_{0}$, T$_{0}$ and $\Lambda$.  The best-fitting values of the GRBs in our sample are shown in Table \ref{table:best_fit}, where we also show the initial spin energy of the magnetar central engine ($E_{\rm p}$) and the spin-down timescale ($t_{\rm d}$).

\section{Results}
\label{sec:Results}

\subsection{GRB 041006}

The plateau duration of GRB~041006 was found to be $T_{0}=18317$~s, which corresponds to a spin-down time (eq. \ref{equ:tp_T_link}) of about 10.1~hr.  We found $B=10.6\times10^{15}$~G and $P=30.6$~ms.  The model fits the early LC reasonably well, but the SN phase is underluminous by a factor of $\Phi = 14.0$, implying that considerably more heating is required to power the SN.  The diffusion timescale was found to be $t_{\rm diff} = 14$~d.  No X-ray data were obtained in order to compare with the plateau duration and luminosity of the optical data.

\subsection{GRB 050525A / SN 2005nc}

We used the decay constants and break times from Blustin et al. (2006) when creating the X-ray luminosity LC.  In the optical regime, we found $T_{0} = 6025$~s, while in the X-rays we found $T_{0} = 720$~s, which are entirely inconsistent.  These durations correspond to spin-down times of 3.4~hr and 0.5~hr, respectively.  From the independent fits, we found $B=10.5\times10^{15}$~G and $P=18.$~ms (optical), and $B=14.5\times10^{15}$~G and $P=8.6$~ms (X-ray).  The diffusion timescale was found to be $t_{\rm diff} = 13.1$~d, and an additional factor of $\Phi = 3.5$ was needed to match the model to the SN luminosity, again implying extra heating is needed to power SN~2005nc.

As a sanity check, we attempted to fit the $R$-band LC using the value of $T_{0}$ determined from the X-ray LC, thus allowing $L_{0}$, $\Lambda$, $t_{\rm diff}$ and $\Phi$ to be free parameters.  However, no acceptable fit was obtained, and the model completely failed to reproduce the SN-phase of the LC.


Finally, we note that the fitted model (eqs. \ref{equ:magnetar_AG} and \ref{equ:SPL}) to the X-ray LC can be considered as somewhat contrived in the sense that the LC is well described by just a SPL with $\alpha=1.52\pm0.03$.  

\subsection{GRB 090618}

We used the decay constants and break times determined by Page et al. (2013).  The plateau duration was found to be 15769~s and 3647~s in the optical and X-ray, respectively, which correspond to spin-down times of 8.9~hr and 0.5~hr.  It is seen that $T_{0}$, and hence the spin-down timescale, is roughly five times larger from modelling of the optical data relative to the X-ray data.  The corresponding $B$-field and spin periods were found to be $B=5.9\times10^{15}$~G and $P=16.3$~ms (optical), and $B=2.9\times10^{15}$~G and $P=3.8$~ms (X-ray).  The diffusion timescale was found to be $t_{\rm diff} = 11.4$~d.  As found for the previous GRB-SNe, additional heating is required to power the accompanying SN, where we find $\Phi = 3.24$.

The model proved to be a very poor fit to the X-ray LC (see Fig. \ref{fig:GRB_Xray_multiplot}), where the later epochs ($t-t_{0}>4000$~s) are more luminous than the model.  Moreover, we were entirely unable to fit the $R$-band LC using the value of $T_{0}$ determined from fitting the X-ray LC, which wholly failed to reproduce the SN luminosity and duration.

\subsection{GRB 091127 / SN 2009nz}

We used the decay constants and break times determined by Troja et al. (2013).  Due to lack of $R$-band data during peak SN light we had to fix the diffusion timescale to 13~d, which was chosen to produce a good visual fit to the data during the fitting process.

We found plateau durations in the optical and X-rays of 16594~s and 14063~s, respectively, which correspond to spin-down timescales of 9.1~hr and 7.9~hr.  The timescales are  consistent between the optical and X-ray regimes.  The corresponding $B$-field and spin periods were found to be $B=5.4\times10^{15}$~G and $P=15.2$~ms (optical), and $B=1.1\times10^{15}$~G and $P=2.9$~ms (X-ray).  Despite the similar plateau durations, the much larger luminosity of the X-ray LC ($\sim 100$ times more luminous than the $R$-band LC) translates to smaller values of $B$ and $P$ due to their $L^{-1/2}_{0}$ dependency.

When using the value of $T_{0}$ from the X-rays to fit the $R$-band LC, a very suitable fit is obtained (which is not surprising given the similar values of the plateau duration at both wavelengths), where we find $B=6.0\times10^{15}$~G and $P=15.7$~ms, which are similar to those obtained when all parameters varied freely. 

While the similar plateau durations found for both the $R$-band and X-ray LCs suggest that a magnetar central engine may be acting in this case, the possible magnetar central engine cannot be responsible for powering SN~2009nz, where we find $\Phi = 3.32$.  Moreover, the fitted model to the X-ray LC can be considered as somewhat contrived and unnecessary as a broken power-law (BPL) also provides an acceptable fit ($T_{\rm B} = 23395.5\pm12747.9$~s, $\alpha_{1} = 0.96\pm0.06$ and $\alpha_{2} = 1.67\pm0.05$), which agrees with the values found by e.g. Filgas et al. (2011) in the optical filters, although we find steeper values for the decay constant before the break, although the time of the break and decay constant thereafter are in agreement within our respective errorbars.

\subsection{ULGRB 111209A / SN 2011kl}

We used the decay slopes and break times from Levan et al. (2014).  The plateau duration was found to be 281201~s and 208336~s in the optical and X-ray\footnote{Data before 12~ks were not included in the X-ray fit.}, respectively, which correspond to spin-down times of 6.5~d and 4.8~d.  The corresponding $B$-field and spin periods were found to be $B=1.1\times10^{15}$~G and $P=13.0$~ms (optical), and $B=1.1\times10^{15}$~G and $P=11.5$~ms (X-ray), which are remarkably consistent.  The diffusion timescale was found to be $t_{\rm diff} = 14.9$~d.  However, unlike the other GRB-SNe considered here, additional heating is not required to power the accompanying SN, where we find $\Phi = 0.73$.  Encouragingly, the initial spin energy of the magnetar central engine was found to be approximately the same from modelling of the two LCs, were we found $E_{\rm p} = $ 1.2 and 1.5 $\times 10^{50}$ erg, respectively.  

Despite the minor disparity of plateau durations between the two frequency regimes, when we use the value of $T_{0}$ determined from the X-ray fitting to the $R$-band data, a very acceptable fit is obtained.  Here we found $t_{\rm diff} = 13.3$~d and $\Phi = 1.23$, and $B=1.3\times10^{15}$~G and $P=12.7$~ms, the latter of which are very similar to that determined from the fit where all parameters varied freely.

Unlike the other events, there is considerable self-consistency seen in the modelling of both the $R$-band and X-ray data.  The plateau durations are entirely consistent with each other, as too are the values of the magnetic field strength and initial spin period, where we find $B=1.1-1.3\times10^{15}$~G and $P=11.5-13.0$~ms.  This arises because the luminosities of the X-ray and $R$-band LCs are more similar than seen for GRB~091127, indeed they are almost identical at times later than $10^{5}$~s.   Moreover, only for ULGRB~111209A / SN~2011kl is the value of $\Phi$ close to unity, which implies that a magnetar central engine can provide enough energy to power all phases of this event, including the prompt emission, AG phase and SN.

\subsection{ULGRB 121027A}

We determined the decay slopes and break times of ULGRB~121027A, finding: $\alpha_{1} = 0.60\pm0.01$, $\alpha_{2} = 5.69\pm0.05$ and $\alpha_{3} = 1.31\pm0.03$, with break times of $T_{\rm B,1} = 16487.6\pm81.3$~s and $T_{\rm B,2} = 520130\pm78631$~s.

At a redshift of $z=1.773$ (Kr\"uehler et al. 2012), the accompanying SN was much too faint to be detected with current technology (e.g. Cano 2013).  As such, the optical data, for which very little exists (e.g. Levan et al. 2014) was not considered here.  Instead we modelled the X-ray LC to see if similar values of $P$ and $B$ would be obtained for this ULGRB relative to the other ULGRB event considered here (111209A).  

In the modelling process, we excluded all data before 9~ks, thus as for ULGRB~111209A, we only fit data obtained during the steep decline phase and the following plateau phase.  We found a plateau duration of $T_{0}=90837$~s, which corresponds to a spin-down time of 2.1~d, which is similar to that found for ULGRB~111209A.  We found $B=0.8\times10^{15}$~G and $P=5.3$~ms

\subsection{GRB 130831A / SN 2013fu}

We determined the decay slopes and break times of GRB~130831A, finding: $\alpha_{1} = 3.31\pm0.22$, $\alpha_{2} = 0.82\pm0.02$, $\alpha_{3} = 4.11\pm1.06$ and $\alpha_{4} = 1.54\pm0.76$, with break times of $T_{\rm B,1} = 210.0\pm60.0$~s, $T_{\rm B,2} = 95536.0\pm19593.4$~s and $T_{\rm B,3} = 121057.5\pm8856.2$~s.

Due to lack of optical data before $t-t_{0} = 3.5$~hr, the early LC evolution is poorly constrained. During the fit, the program was not able to fit the data very well, and only approximate values, determined by trial-and-error of numerous fitting iterations, were determined for the  free parameters in the model.

We found (approximate) plateau durations in the optical and X-rays of 5570~s and 3208~s, respectively, which correspond to spin-down timescales of 3.1~hr and 1.7~hr.  The corresponding $B$-field and spin periods were found to be $B=12.9\times10^{15}$~G and $P=21.3$~ms (optical), and $B=7.3\times10^{15}$~G and $P=9.2$~ms (X-ray).  We found a diffusion timescale of $t_{\rm diff} \approx 12$~d and $\Phi \approx 6.4$, the latter indicating that additional sources of heating are needed to power SN~2013fu.

Moreover, when using the value of $T_{0}$ from the X-rays when fitting the $R$-band data, we found $t_{\rm diff} \approx 14.0$~d and $\Phi \approx 7.2$, which are consistent with the previous fit, and which still indicate the needed for more heating during the SN phase.  The resultant values of the initial spin period and magnetic field strength were $P=16.2$~ms and $B=14.1\times10^{15}$~G, respectively.

We note that the fitted model proved to be a very poor fit to the X-ray data (see Fig. \ref{fig:GRB_Xray_multiplot}).  If this model is correct, it require two additional flaring episodes at $t-t_{0} = 20-100$ ks, and $5\times10^5$~s and $2\times10^6$~s.  The fitted model in the X-rays is not overly convincing, and it can be argued that the $R$-band model is as equally unconvincing, especially when we consider that the model has artificially created a plateau phase that is entirely unsupported by observations.

\section{Discussion}
\label{sec:discussion}

\subsection{What powers the supernova luminosity?}

In the previous section we presented the results of fitting our combined magnetar model to the optical and X-ray LCs of five LGRBs and their accompanying SNe, and two ULGRBs, one of which has an accompanying SN (ULGRB~111209A / SN~2011kl).  For each of the LGRB-SN events, it was seen that while the model can be well fit to the early AG-dominated phase, the predicted SN luminosity was too faint compared with observations.  Indeed in the most extreme case of GRB~041006, the predicted magnetar-powered SN was more than 17 times fainter than the observations.  For the remaining LGRB-SN events, the magnetar-powered SN was underluminous by a factor of 3--17, with three of the five events clustering around $\Phi \approx 3.0-3.5$.  

In contrast, when fitting the optical data of ULGRB~111209A / SN~2011kl, and letting all of the parameters vary freely, it was seen that the model slightly over-estimated the SN luminosity, where a factor of $\Phi=0.73$ was needed to get the model to match the observations.  At first glance, this might suggest some difference in the efficiencies between the AG-dominated and SN-dominated phases in this event.  However, when the optical LC was fit with the plateau duration of the SPL component determined from the X-ray data, it was seen that the SN luminosity was slightly under-predicted, where a factor of $\Phi=1.23$ was determined.  Given the assumptions that have gone into the analytical models, and the inherent uncertainties that subsequently arise, we can conclude that both scenarios predict $\Phi\approx1$, and do so in a very self-consistent manner.  This also suggests that the efficiencies required during the AG- and SN-dominated phases are approximately the same (within 30\%).

The apparent discrepancy between the magnetar-powered AG phase and the magnetar-powered SN phase of the GRB-SNe proves strong evidence that additional sources of heating are needed to power the associated SNe.  Let us consider the various energy sources that can drive and power a SN explosion: (1) radioactivity, (2) shock-generated energy arising from the collision of SN ejecta with circumstellar material, and (3) a magnetar-powered SN.  In the case of GRB-SNe, it is possible that some shock-interaction energy could be powering some portion of the SN output during the peak photospheric phase (e.g. Fryer et al. 2007).  Moreover, it is also possible that the initial bumps observed for SN~2006aj (associated with GRB~060218; Campana et al. 2006; Pian et al. 2006; Ferrero et al. 2006) and SN~2010bh (associated with GRB~100316D; Starling et al. 2011; Cano et al. 2011b; Bufano et al. 2012; Olivares et al. 2012) are powered by material heated by the passage of the shock-breakout through the SN ejecta.  

In this paper we investigated the possibility that GRB-SNe could be powered solely by rotationally extracted energy from a magnetar central engine.  The values of $\Phi$ seem to indicate that this is very unlikely.  As such, the main power source of GRB-SNe is likely to be energy input from the radioactive decay of nickel into cobalt, and cobalt into iron.  Such a proposition is entirely in line with countless previous GRB-SN investigations (see e.g. Woosley \& Bloom 2006; Hjorth \& Bloom 2013; Cano 2013), where the inferred nickel masses of GRB-SNe range from 0.1--0.6~M$_{\odot}$, where the lower limit corresponds to SN~2006aj (Pian et al. 2006; Mazzali et al. 2006a) and SN~2010bh (Cano et al. 2011b; Cano 2013), and the upper limit corresponds to SN~2012bz (Schulze et al. 2014; Cano 2013) and SN~2003lw (Malesani et al. 2004; Mazzali et al. 2006b).  Moreover, this range of nickel masses is roughly twice that inferred for SNe Ibc ($\approx 0.2$~M$_{\odot}$; Drout et al. 2011; Cano 2013; Lyman et al. 2014; Taddia et al. 2015), which are also thought to be powered by radioactive heating.

\subsection{ULGRBs versus LGRBs with associated SNe}

In contrast to the scenario discussion in the previous section, we have shown in this work that emission from a magnetar central engine \emph{can} can power all phases of an ULGRB event, from the AG-dominated phase to the SN-dominated phase.  By considering different permutations of fitting our analytical model to the optical and X-ray LCs of ULGRB~111209A, we constrained a magnetic field strength of $B = 1.1-1.3\times10^{15}$~G, an initial spin period of $P=11.5-13.0$~ms, and a spin-down time of $t_{\rm d} = 4.8-6.5$~d.  The initial energy of the magnetar central engine was found to be $E_{\rm p} = 1.2-1.6\times10^{50}$ erg, with excellent agreement derived from modelling both the optical and X-ray LCs.  Additionally, for ULGRB~121027A, from modelling of the X-ray LC we determined $B=0.8\times10^{15}$~G, $P=5.3$~ms, a spin-down time of 2.1~d and $E_{\rm p} = 7.2\times10^{50}$ erg.  There is very good agreement in these values between the two ULGRB events.


Turning to the literature, a few authors have considered energy arising from a magnetar central engine for ULGRB~111209A / SN~2011kl.  Greiner et al. (2015) showed compelling evidence that the peak SN light could not be powered entirely (or at all) by radioactive heating.  Their argument was based primarily on the fact that the inferred ejecta mass ($3.2\pm0.5$~M$_{\odot}$), determined via fitting the Arnett (1982) model to their constructed bolometric LC, was too low for the amount of nickel needed to explain the observed bolometric luminosity ($1.0\pm0.1$~M$_{\odot}$).  The ratio of $\frac{\rm M_{Ni}}{\rm M_{ej}} = 0.3$ was much larger than that inferred for the general GRB-SN population ($\frac{\rm M_{Ni}}{\rm M_{ej}} \approx 0.07$; Cano 2013), which rules against radioactive heating powering SN~2011kl.  

Indeed, their more convincing evidence was the shape and relative brightness of an optical spectrum obtained of SN~2011kl just after peak SN light ($t-t_{0} = 20$~d, rest-frame), which was entirely unlike the spectra observed for GRB-SNe, including SN~1998bw (Patat et al. 2001).  Instead, the spectrum more closely resembled those of SLSNe in its shape, including the sharp cutoff at wavelengths bluewards of 3000~\AA.  As the likely energy source of many SLSNe-I (type I) is expected to be energy injection from a magnetar central engine (e.g. Chatzopoulos et al. 2011; Inserra et al. 2013; Nicholl et al. 2013), logic dictates that SN~2011kl could also be powered by similar means.  

In the model used by Greiner et al. (2015) to create synthetic spectra from a magnetar central engine, the input values that best reproduced the observed spectrum were $P=12$~ms and $B=0.6-0.9\times10^{15}$~G.  They also assumed an explosion energy of $5.5\times10^{51}$~erg and a grey opacity of 0.07~cm$^{2}$~g$^{-1}$. The values of $P$ and $B$ are in excellent agreement with those found here.  Using eq. \ref{equ:mag_SN_B}, this implies spin-down times from 8.4--19.0~d (compared with 4.8--6.5~d found here).

Next, M15 also considered a magnetar central engine that could power both the initial prompt emission and the subsequent SN associated with ULGRB~111209A.  In their model they assumed an ejecta mass of 3~M$_{\odot}$, a nickel mass of $M_{\rm ni} = 0.2$~M$_{\odot}$, and then constrained the initial spin period to be $P\approx2$~ms and $B\approx 0.4\times10^{15}$~G.  In their model they assume that the spin-down time is approximately equal to the prompt emission duration, where they constrain a value of $\approx$14~ks, which is about 3.9~hr (i.e. an order of magnitude smaller than the spin-down timescale than determined in our model).  We note that using our eq. \ref{equ:mag_SN_B}, these values imply a spin-down time of $\approx 1.2$~d.  

Additionally, Bersten et al. (2016; B16 hereafter) used a one-dimensional LTE radiation hydrodynamics code to model the bolometric LC of SN~2011kl.  In their model they assumed an ejecta mass of 2.5~M$_{\odot}$, a nickel mass of 0.2~M$_{\odot}$ and an explosion energy of $5.5\times10^{51}$~erg.  Just as M15, they assumed an opacity of $\kappa = 0.2$~cm$^{2}$~g$^{-1}$.  The preferred model of B16 was for $P=3.5$~ms and $B=1.95\times10^{15}$~G, which implies a spin-down time of 0.15~d (eq. \ref{equ:mag_SN_B}).  Here the spin period is commensurate with that determined by M15, although their magnetic field strength is five times larger than M15's, but quite similar to that found here.  

The results of B16 are quite enlightening, and show how the effect of using different values of the grey opacity lead to different derived values of $P$: in Fig. 1 of B16, the results of Greiner et al. (2015) also provide a very acceptable fit to the bolometric LC of SN~2011kl, and moreover, go someway towards refuting the assertion of B16 that additional nickel is needed to power the late-time bolometric LC.  Thus, different values of the grey opacity used in each of the studies leads to a degeneracy in the derived values of $P$ and $B$, but particularly $P$: it is currently impossible to determine between $P\approx 2$~ms and $P\approx12$~ms.  Larger values of $\kappa$ inevitably lead to larger diffusion timescales, which makes the derived value of $P$ less certain.  However, in our modelling so such degeneracy is present -- all we constrain is the diffusion timescale and not the ejecta mass.  For the sake of comparison, we can use the diffusion timescale approximation (Arnett 1979; Chatzopoulos et al. 2013) to estimate the ejecta mass of SN~2011kl: 

\begin{equation}
M_{\rm ej} = \frac{3}{10} \frac{\beta c}{\kappa} v_{\rm ph} t_{\rm diff}^{2}
\end{equation}

Using the ejecta velocity determined by Greiner et al. (2015), $v_{\rm ph} = 21,000$~km~s${^-1}$, $\kappa = 0.07$~cm$^{2}$~g$^{-1}$, $\beta \approx 13.8$, and $t_{\rm diff} = 14.92$~d (Table \ref{table:best_fit}), we find an ejecta mass of $\approx 3.1$~M$_{\odot}$, which is nearly identical to that used/found in the other analyses.

When comparing the inferred properties of the magnetar central engines of ULGRBs to those of LGRB-SNe ($P$, $B$ and the spin-down timescale), if we assume that both sets of events are powered by EM radiation from a magnetar central engine, we can draw the following general conclusions.  First, the spin-down timescales of the ULGRBs are roughly an order of magnitude longer than those of the LGRB-SNe (a couple of days versus a few to tens of hours, respectively).  This is in agreement with the idea that the luminosity of a magnetar-powered SN is less dependent on the total energetics available, but rather on how long the central engine is active, where central engines with longer durations give rise to brighter SNe, as modelled here for SN~2011kl relative to all other LGRB-SNe.  This idea is also in keeping when considering SN~2011kl relative to SLSNe, where the spin-down timescales of SLSNe are of order months to years (e.g. Nicholl et al. 2013), thus providing a natural explanation why SN~2011kl was not as luminous as SLSNe.   Secondly, the magnetic field strengths of the two ULGRBs are, on average, less than those inferred for the LGRB-SNe.  This general result, which is not statistically significant, is in general keeping with the notion that magnetars with weaker $\vec{B}$-fields will produce prompt emission that is weaker and longer lasting.  Finally, the initial spin periods of the two ULGRBs are, on average, less than those implied for the GRB-SNe, and commensurate with those inferred for SLSNe.  

\subsection{Model Diagnostics}

It could be argued that the inferred $P$ and $B$ values of the LGRB-SNe are not very informative in the sense that a magnetar central engine may not actually be present in these events.  Instead, an accreting BH central engine could be driving these LGRB-SN events.  In this scenario, the energetics and duration of the prompt emission are directly related to the initial spin of the BH and the accretion rate, and the stellar envelope infall time, respectively.  The SN would then be powered by radioactive heating, where heavy elements, including nickel, are nucleosynthesised by the energetic wind emitted by the accretion disk (e.g. MacFadyen \& Woosley 1999; Nagataki et al. 2006).

In an attempt to drawn a consistent picture of ULGRB-SNe vs. LGRB-SNe, we conclude that there are three possible scenarios of how the luminosity of an LGRB-SN is powered:

\begin{enumerate}
 \item Magnetar central engines are only present for ULGRBs, while those in LGRB-SNe are accreting BHs, where the SN is powered only by radioactivity.
 \item Magnetar central engines may be present for LGRB-SNe events, and the SN is powered by a combination of magnetar emission AND radioactivity.
 \item Magnetar central engines may be present for LGRB-SNe events, and the SN is powered by just radioactivity.
\end{enumerate}

However, the second scenario above can probably be ruled out for a couple of reasons.  First is the general shape of GRB-SN spectra, e.g. SN~1998bw, SN~2010bh, etc., relative to SN~2011kl, where the shape of the latter (Fig. 3 in Greiner et al. 2015) is entirely unlike those of the former.  While all spectra are lacking H and He features (and thus their type Ic classification), the spectrum of SN~2011kl is featureless redwards of 3000~\AA, and the spectral undulations that are typical of SNe IcBL are clearly lacking.  Intriguingly, the flux of the spectrum of SN~2011kl does not diminish in the $3000-4000$~\AA~region, as seen for e.g. SN~1998bw.  Instead, the general shape of the spectrum of SN~2011kl is reminiscent of those of SLSNe, including the sharp cut-off at wavelengths bluewards of 3000~\AA, and the featureless regime redwards to 5000~\AA~(where the spectrum ends).

The second and third scenarios can be doubted when we consider whether the collapsar and magnetar models can generate enough nickel to describe the observed luminosities of the LGRB-SNe in our sample.  In both explosion models, a small amount of nickel can be explosively nucleosynthesised if a large amount of energy is deposited into a relatively compact volume in a short period of time.  For an explosion energy of $10^{52}$~erg that is focused into $\sim1$\% of the star, and which occurs in the region between the newly formed compact object and $4\times10^9$~cm (MacFadyen et al. 2001), temperatures of $5\times10^9$~K can be attained, leading to roughly 0.1~M$_{\odot}$ of nickel being nucleosynthesised.  However, the precise amount of $^{56}$Ni that is generated is quite uncertain, and depends greatly on how much the star has expanded (or collapsed), prior to energy deposition.  

Moreover, it is difficult to produce a sufficient amount of $^{56}$Ni via energy injection from a central engine rather than a prompt explosion in the explosive nucleosynthesis scenario (e.g. Maeda \& Tominaga 2009).  For example, simulations suggest that only a a few hundredths of a solar mass of nickel can be synthesized in the magnetar model (e.g. Barkov \& Komissarov 2011).  However it may be possible to generate more nickel either by tapping into the initial rotational energy of the magnetar via magnetic stresses, thus enhancing the 
shocks induced by the collision of the energetic wind emanated by the magnetar with material already processed by the SN shock (Thompson 2007; Thompson et al. 2010).  However, in this scenario an isotropic-equivalent energy input rate of more than 10$^{52}$~erg is required, and the subsequent procurement of additional nickel mass via explosive nucleosynthesis will inevitably lead to a more rapid spin-down of the magnetar central engine, rendering it unable to produce energy input during the AG phase.  However, as we have shown here, such energy injection may not be required.  It is also worth considering that if a magnetar (and the subsequent GRB) is formed via the accretion-induced collapse of a white dwarf star, or perhaps the merger of two white dwarfs, there is no explosive nucleosynthesis and thus a very low $^{56}$Ni yield (e.g. Metzger et al. 2007).  



In the collapsar model, there are additional physical processes that can lead to the creation of greater masses of radioactive nickel.  Another potential source of $^{56}$Ni arises from the wind emitted by the accretion disk surrounding the newly formed BH.  According to the numerical simulations of MacFadyen \& Woosley (1999), the amount of generated nickel depends on the accretion rate as well as the viscosity of the inflow.  In theory, at least, the only upper bound on the amount of nickel that can be synthesized by the disk wind is the mass of material that is accreted.  In an analytical approach, Milosavljevi\'c et al. (2012) demonstrated that enough $^{56}$Ni can be synthesized (in order to match observations of GRB-SNe), over the course of a few tens of seconds, in the convective accretion flow arising from the initial circularization of the infalling envelope around the BH.

Thus, in the collapsar model, there are more physical processes occurring to generate the needed required nickel masses to reproduce the observed luminosities of GRB-SNe.   We acknowledge that it may also be possible to generate the required nickel mass in the magnetar scenario, although the physics behind its production are admittedly less well understood.  While it may be possible to explosively synthesise up to a tenth of a solar mass of nickel in the magnetar scenario, enough to account for the observed luminosities of SN~2006aj and SN~2010bh, this is not nearly enough to explain the observed luminosities of other GRB-SNe such as SN~1998bw, SN~2003dh, SN~2003lw, SN~2012bz and SN~2013cq.  Instead, only the collapsar model offers the physical means to generate enough nickel to explain the observed luminosities of these events, leading us to conclude that GRB-SNe are likely powered entirely by radioactive heating produced by collapsars, and not magnetar central engines.


\subsection{Physical Scenario}

The model considered here is in many ways analogous to that considered by M15.  Both models considered a scenario where at very early times ($<<$ spin-down timescale) the jet cleanly escapes the star, and a large fraction of the rotationally extracted energy goes towards powering the prompt emission.  For a (narrow) jet to be formed, the magnetar outflow is collimated via confinement from the stellar envelope (Uzdensky \& MacFadyen 2006; Bucciantini et al. 2007).  At later times ($>>$ spin-down timescale) the jet becomes trapped such that the remaining energy thermalises behind the ejecta with a high enough efficiency to then power the SN.  M15 suggest that such a scenario may be physically viable if at late times the jet becomes less stable, or less effective at maintaining an open cavity through the expanding SN ejecta.

\subsection{Caveats}

In our analytical approach we have made several assumptions, including a canonical NS mass (1.4~M$_{\odot}$) and radius (10~km), and 100\% trapping of magnetar-driven EM emission into the AG and SN.

Perhaps the biggest assumption underpinning our model is that the energy powering an entire LGRB/ULGRB event is equally divided between the AG and SN components.  Instead, it is possible that different fractions of the total energy will go to power each component of a given event, where this fraction is currently unknown.  

Let us first consider LGRBs vs. low-luminosity GRBs ($ll$GRBs), where the former are expected to be jetted-GRB events.  Many authors (Soderberg et al. 2006; Margutti et al. 2013; Margutti et al. 2014; Cano et al. 2015) have demonstrated that the amount of kinetic energy contained in the mildly-relativistic (non-thermal) component in $ll$GRBs is roughly 1\% (perhaps less) of the kinetic energy contained in the SN component in these events.  In contrast, the kinetic energy contained in the relativistic component in LGRB events is of order  that associated with the SN component.  Therefore, in the context of our model, one would expect large values of $\Phi$ if we modelled the AG and SN components of any $ll$GRBs.  However, we have not modelled any $ll$GRBs in this study, only LGRBs and ULGRBs.

Instead, the LGRBs/ULGRBs considered here are all thought to arise from jet-driven explosions, which are more collimated than $ll$GRB explosions, and it is expected that there is more energy contained in the central engine driving the explosions.  If the radiated energy modelled here is a good proxy for the total kinetic energy associated with each component of the explosion, then we would expect that similar amounts of energy will be associated with the AG and SN in jetted events.  Within this framework, we cannot envision a scenario where there should be a difference in the fraction of energy imparted to the AG and SN phases of a LGRB relative to an ULGRB if both arise from jetted events.  If such a difference does exist, then the $\Phi$ factor derived here could be used as an indicator of the relative efficiencies powering the AG and SN phases in an LGRB event.  However, one would then need to understand and explain why for ULGRBs the partitioning of energy between the AG and SN is nearly 50/50 (as indicated by our result that $\Phi\approx1$), and why this not the case for LGRBs.

\section{Conclusions}
\label{sec:conclusions}

In this paper we presented an analytical model that considers energy arising from a magnetar central engine.  The motivation for deriving this model was to test the hypothesis that a  magnetar could power an entire LGRB/ULGRB event (prompt emission, AG and SN).  The results of modelling five LGRBs and two ULGRBs, including their associated SNe, show that a magnetar central engine cannot be solely responsible for producing the observed luminosity of an LGRB-SN.  When we fit our model to the optical and X-ray LCs of the LGRB-SNe in our sample, while the early AG-dominated phase can be reasonably well described with our model, the predicted SN luminosity in each event is underluminous by a factor of 3--17.  We use this as strong evidence that additional sources of heating must be present to power the observed luminosity of LGRB-SNe, which is very likely to be heating arising from the thermalisation of $\gamma$-rays emitted during the radioactive decay of nickel into cobalt, and cobalt into iron.

However, our magnetar model was able to successfully describe all phases of ULGRB~111209A / SN~2011kl, from the early AG to the later SN.  Moreover, no matter whether we use the best-fitting values of the free parameters in our model derived from just the optical LC, or whether we use the plateau duration derived from modelling the X-ray LC when fitting the optical LC, we obtain the same result.  This self-consistent approach implies that the magnetar central engine of ULGRB~111209A/ SN~2011kl had a magnetic field strength of $B = 1.1-1.3\times10^{15}$~G, an initial spin period of $P=11.5-13.0$~ms, a spin-down time of $t_{\rm d} = 4.8-6.5$~d and an initial energy of $E_{\rm p} = 1.2-1.6\times10^{50}$~erg, which are entirely consistent with those determined by other authors.  Encouragingly, the close similarities in the values derived from both the optical and X-ray data provide credence to our modelling approach, which strongly suggests that the luminosity of ULGRB~111209A and SN~2011kl was powered entirely by emission from a magnetar central engine.  Finally, we also modelled the X-ray LC of ULGRB~121027A, finding $B=0.8\times10^{15}$~G, $P=5.3$~ms and a spin-down time of 2.1~d.  These values for the two ULGRBs are in good agreement with each other. 

To place our results in the context of SLSNe, a natural way to explain why SN~2011kl was less luminous than SLSNe (which are also thought to be powered by emission from a magnetar central engine) is if we consider the spin-down timescales of both sets of events.   The spin-down timescales of SLSNe are of order months to years (e.g. Nicholl et al. 2014), compared with a few days for SN~2011kl.  This measurement is in agreement with the idea that the luminosity of a magnetar-powered SN is directly related to how long the central engine is active, where central engines with longer durations give rise to brighter SNe. 

Next, we concluded that GRB-SNe are not powered by emission arising from a magnetar central engine, but instead could be powered by (1) either solely from radioactive heating (and thus the central engine is likely not a magnetar but rather an accreting BH), or (2) a combination of radioactivity and energy from a magnetar central engine.  This second scenario can likely be ruled out based on the shape of GRB-SN spectra, which are entirely unlike those of SLSNe, and via results from simulations that show that very little nickel is synthesized in the magnetar model.

Finally, further investigations are needed to determine if ULGRB-SNe are standardizable candles in the same manner as GRB-SNe (Cano 2014) and SLSNe (Inserra \& Smartt 2014), and either follow the luminosity--decline relationship seen for GRB-SNe (Cano \& Jakobsson 2014; Li \& Hjorth 2014), or perhaps follow their own luminosity--decline relationship.  This latter hypothesis can only be verified with the detection of future ULGRB-SNe.

\section*{Acknowledgments}

We wish to thank Palli Jakobsson and Gulli Bjornsson for helpful comments of the original manuscript.  We also acknowledge and thank the anonymous referee for their constructive discussion of the submitted paper.  The work of ZC was funded both by a short-term fellowship awarded by the Japan Society for the Promotion of Science (JSPS), and from a Project Grant from the Icelandic Research Fund.  AJ is funded by the University of Iceland's Research Fund for Doctoral Students.  The work by K.M. is partly supported by JSPS KAKENHI Grant 26800100, and by World Premier International Research Center Initiative (WPI Initiative), MEXT, Japan.

\begin{landscape}
\thispagestyle{empty}
\begin{table}
\centering
\setlength{\tabcolsep}{5pt}
\setlength{\extrarowheight}{5pt}
\caption{Best-fitting values of the fitted magnetar model}
\label{table:best_fit}
\begin{tabular}{cccccccccccc}
\hline																													
GRB	&		&	$B$ (10$^{15}$ G)	&	$P$ (ms)	&			$L_{0}$	 (erg~s$^{-1}$)						&		$T_{0}$	(s)			&		$E_{\rm p} = L_{0}T_{0}$ (10$^{50}$~erg)				&			$t_{\rm p}$ (d)			&			$\Lambda$	 (erg~s$^{-1}$)						&	$\alpha = \Gamma^{\dagger}_{\gamma} + 1$	&		$t_{\rm diff}$ (d)				&		$\Phi$				\\
\hline																																																											
041006	&	Optical	&	10.6	&	30.6	&	$	(	1.18	\pm	0.06	)	\times	10^{45}	$	&	$	18317.3	\pm	693.0	$	&	$	0.22	\pm	0.01	$	&	$	0.42	\pm	0.02	$	&	$	(	6.52	\pm	0.30	)	\times	10^{52}	$	&	2.37	&	$	13.95	\pm	0.75	$	&	$	13.95	\pm	0.69	$	\\
041006	&	X-ray	&	-	&	-	&			-							&		-				&						&						&			-							&	-	&		-				&		-				\\
050525A	&	Optical	&	10.5	&	18.0	&	$	(	1.05	\pm	0.05	)	\times	10^{46}	$	&	$	6025.1	\pm	235.6	$	&	$	0.63	\pm	0.02	$	&	$	0.14	\pm	0.01	$	&	$	(	1.67	\pm	0.17	)	\times	10^{51}	$	&	2.00	&	$	13.10	\pm	1.07	$	&	$	3.47	\pm	0.27	$	\\
050525A	&	X-ray	&	14.5	&	8.6	&	$	(	3.85	\pm	0.18	)	\times	10^{47}	$	&	$	720.1	\pm	34.2	$	&	$	2.77	\pm	0.03	$	&	$	0.02	\pm	0.00	$	&	$	(	3.87	\pm	0.19	)	\times	10^{51}	$	&	2.00	&		-				&		-				\\
090618	&	Optical	&	5.9	&	16.3	&	$	(	4.89	\pm	0.06	)	\times	10^{45}	$	&	$	15768.9	\pm	152.7	$	&	$	0.77	\pm	0.01	$	&	$	0.37	\pm	0.00	$	&	$	(	5.09	\pm	0.17	)	\times	10^{52}	$	&	2.42	&	$	11.42	\pm	0.45	$	&	$	3.24	\pm	0.18	$	\\
090618	&	X-ray	&	2.9	&	3.8	&	$	(	3.80	\pm	0.05	)	\times	10^{47}	$	&	$	3647.4	\pm	46.0	$	&	$	13.86	\pm	0.05	$	&	$	0.08	\pm	0.00	$	&	$	(	2.69	\pm	0.03	)	\times	10^{53}	$	&	2.42	&		-				&		-				\\
091127	&	Optical	&	5.4	&	15.2	&	$	(	5.31	\pm	0.02	)	\times	10^{45}	$	&	$	16593.7	\pm	60.8	$	&	$	0.88	\pm	0.01	$	&	$	0.38	\pm	0.00	$	&	$	(	1.42	\pm	0.03	)	\times	10^{52}	$	&	3.05	&	$	13^{*}			$	&	$	3.32	\pm	0.17	$	\\
091127	&	X-ray	&	1.1	&	2.9	&	$	(	1.72	\pm	0.12	)	\times	10^{47}	$	&	$	14063.4	\pm	696.4	$	&	$	24.19	\pm	0.70	$	&	$	0.33	\pm	0.02	$	&	$	(	3.57	\pm	0.17	)	\times	10^{57}	$	&	3.05	&		-				&		-				\\
111209A	&	Optical	&	1.1	&	13.0	&	$	(	4.34	\pm	0.48	)	\times	10^{44}	$	&	$	281201.1	\pm	28874.9	$	&	$	1.22	\pm	0.03	$	&	$	6.51	\pm	0.67	$	&	$	(	6.76	\pm	0.37	)	\times	10^{56}	$	&	2.48	&	$	14.92	\pm	1.50	$	&		\textbf{0.73}	$\pm$	\textbf{0.12}		\\
111209A	&	X-ray	&	1.3	&	11.5	&	$	(	7.44	\pm	1.02	)	\times	10^{44}	$	&	$	208336.3	\pm	25931.0	$	&	$	1.55	\pm	0.03	$	&	$	4.82	\pm	0.60	$	&	$	(	4.06	\pm	0.13	)	\times	10^{56}	$	&	2.48	&		-				&		-				\\
121027A	&	Optical	&	-	&	-	&			-							&		-				&						&						&			-							&	-	&		-				&		-				\\
121027A	&	X-ray	&	0.8	&	5.3	&	$	(	7.91	\pm	0.75	)	\times	10^{45}	$	&	$	90837.3	\pm	7364.9	$	&	$	7.19	\pm	0.07	$	&	$	2.10	\pm	0.17	$	&	$	(	3.70	\pm	0.24	)	\times	10^{57}	$	&	2.82	&		-				&		-				\\
130831A	&	Optical	&	12.9	&	21.3	&	$	\approx	8.1				\times	10^{45}	$	&	$	\approx	5570		$	&	$		\approx	0.45	$	&	$	\approx	0.13		$	&	$	\approx	4.6				\times	10^{54}	$	&	2.93	&	$	\approx	12.0		$	&	$	\approx	6.4		$	\\
130831A	&	X-ray	&	7.3	&	9.2	&	$	(	7.62	\pm	0.55	)	\times	10^{46}	$	&	$	3208.4	\pm	179.9	$	&	$	2.44	\pm	0.02	$	&	$	0.07	\pm	0.00	$	&	$	(	1.11	\pm	0.07	)	\times	10^{53}	$	&	2.93	&		-				&		-				\\
\hline							
\end{tabular}
\begin{flushleft}
$^{\dagger}$ $\gamma$-ray photon index obtained from \url{http://swift.gsfc.nasa.gov/archive/grb_table/} for all GRBs, except GRB~041006, which is from the \emph{HETE} website: \url{http://space.mit.edu/HETE/Bursts/GRB041006/}.
\end{flushleft}
\end{table}
\end{landscape}

\bsp

\label{lastpage}

\end{document}